\documentclass[12pt]{article}
\usepackage{graphicx}
\usepackage{amsmath}
\usepackage{amssymb}
\usepackage{booktabs}

\newcommand{\bea}{\begin{eqnarray}}
\newcommand{\eea}{\end{eqnarray}}
\newcommand{\be}{\begin{equation}}
\def\bel#1{\begin{equation} \label{#1}}
\newcommand{\ee}{\end{equation}}
\newcommand{\ba}{\begin{align}}
\newcommand{\ea}{\end{align}}
\newcommand{\comments}[1]{}

\begin{document}
\begin{titlepage}
\begin{flushright}
\parbox[t]{1.4in}{DAMTP-2014-26}
\end{flushright}

\begin{center}

\vspace*{ 1.2cm}

{\Large \bf Local String Models and Moduli Stabilisation}

\vskip 1.2cm

\renewcommand{\thefootnote}{}
\begin{center}
Fernando Quevedo
 \end{center}
\vskip .2cm
\renewcommand{\thefootnote}{\arabic{footnote}}

{\it \small $^1$ ICTP, Strada Costiera 11, Trieste 34014, Italy.\\
$^2$ DAMTP, University of Cambridge, Wilberforce Road, Cambridge, CB3 0WA, UK.\\[0.5cm]}



\end{center}


\begin{center} {\bf ABSTRACT } \end{center}
A brief overview is presented of the  progress made during the past  few years on the general structure of local models of particle physics from string theory including: moduli stabilisation, supersymmetry breaking, global embedding in compact Calabi-Yau compactifications and potential cosmological implications. Type IIB D-brane constructions and  the Large Volume Scenario (LVS)
are discussed in some detail emphasising the recent achievements and the main open questions.
\end{titlepage}
\tableofcontents
\section{Introduction}\label{ra_sec1}

The aim of string phenomenology is well defined and very ambitious: to uncover  string theory scenarios that satisfy all particle physics and cosmological observations and hopefully lead to measurable predictions (for a comprehensive treatment of the field with a very complete set of references   see \cite{iu}).

This defines a list of concrete challenges for string constructions that have been addressed over the years:
\begin{enumerate}

\item{Gauge and matter structure of the Standard Model (SM).}
\item{ Hierarchy of scales and quark/lepton masses (including a proper account of neutrino masses).}
\item{ Realistic flavour structure with right quark (CKM),  lepton (PMNS) mixings and right amount of CP violation,  avoiding  flavour changing neutral currents (FCNC).}
\item{ Hierarchy of gauge couplings at low energies potentially unified at high energy.}
\item{ Almost stable proton but  with a  realistic quantitative account of baryogenesis.}
\item{ Inflation or alternative early universe scenarios that can explain the CMB fluctuations.}
\item{ Dark matter (but avoid overclosing).}
\item{ Dark radiation ($4\geq N_{eff}\geq 3.04$).}
\item{ Dark energy (with equation of state $w=\rho/p\sim -1$).}

\end{enumerate}

Addressing all of these issues has kept the string phenomenology community busy for several decades now. Partial success has been achieved on each of the points but not for all of them at the same time in particular classes of models. This list can be seen as a guideline for the present overview. Notice that in string theory, contrary to field theoretical model building, if a model fails with one of the above requirements it has to be ruled out.

The prospect of obtaining a proper ultraviolet complete extension of the Standard Model (SM) not only justifies efforts in this direction but provides for the first time a well defined alternative to the traditional bottom-up approach to model building beyond the SM that has very limited guidelines beyond experiment and that is currently being under pressure by the LHC results so far.

In order to address these issues, several approaches have been followed. 
Ideally the first attempts are to try as much as possible to extract
`genericÕ model independent implications of string theory. 
Regarding the general string predictions relevant for our universe (see for instance the  discussion in \cite{witten}) we can mention only very few :
\begin{itemize}
\item{} Gravity + dilaton + antisymmetric tensors + gauge fields + matter.

\item{} Supersymmetry (SUSY) (with 32, 16 or less supercharges, but breaking scale not fixed).

\item{} Extra dimensions (6 or 7)
    (flat, small, large or warped).
 
\item{}
No (massless) continuous spin representations (CSR) in perturbative string theory. 
\end{itemize}

These have to be compared with general model independent field theoretical  predictions which are also very few (identity of particles, existence of antiparticles, 
relation between spin statistics, the CPT theorem and the running of physical couplings with energy, following the renormalisation group (RG) equations).  The 4th prediction above is relatively less known \cite{fqt}, it essentially states the fact that in perturbative string theory massive and massless representations of the Poincare group are linked to each other and since the massive representations are finite dimensional the same should be true for the massless representations, forbidding then the continuous spin representations \footnote{Continuous spin representations (CSR) are representations of the little group for massless states of the Poincare group which is the Euclidean group in two dimensions. Eliminating these infinite dimensional unitary representations (arguing for instance that these particles have not been observed in nature \cite{weinberg}) limits to  the subgroup $SO(2)$  with the standard helicity quantum number.}. Notice that being massless they could have been relevant at low energies and perturbative string theory indicates that they should not exist at least as perturbative string states, an statement consistent with all observations that has no clear explanation otherwise.

Having stated the general properties of string models we may concentrate on their general 4-dimensional implications.  The most promising compactifications have $\mathcal{N}=1$ SUSY that guarantees stability and chirality in the spectrum. The generic properties of 4D string compactifications are:
\begin{enumerate}
\item{}Moduli: gravitationally-coupled scalar  fields that usually measure size and shape of extra dimensions. They are massless as long as supersymmetry is unbroken.

\item{} Antisymmetric tensors of different ranks implying the generic existence of axions, the possibility of turning on their fluxes in the extra dimensions and the  generic appearance of 
     branes that couple to these antisymmetric tensors and may even host the Standard Model.
     
\item{} Absence of continuous global symmetries outside those Peccei-Quinn symmetries associated to antisymmetric tensors. Contrary to standard field theories (without gravity) string theory does not allow exact continuous global symmetries \cite{banksdixon}.

\item{} Matter appears on low dimensional group representations:
     (bifundamentals, symmetric, antisymmetric, adjoints).
  
\item{} If the 4D theory has  SUSY broken at the  TeV scale the moduli tend to receive a mass of an order similar to the soft terms, implying the Cosmological Moduli Problem (CMP)
     ( they overclose the universe or ruin nucleosynthesis upon late decay unless the  mass of all the moduli can be made $m>10$ TeV).
\end{enumerate}

These generic predictions are very powerful and can be used to study general `string inspired' scenarios to try to make contact with observations but are clearly not enough for a proper phenomenology and we have to consider concrete ways to explicitly build models. Over the years, two general classes of models have been studied.
\begin{itemize}
\item{} {\it Global string models}: 10D string theory compactified on 6D manifold. Gauge and matter fields in 4D come from gauge multiplets in 10D.
\item{} {\it Local string models}: Standard model lives on a D-brane localised in some point in the extra dimensions.
\end{itemize}

The global models are essentially the heterotic models in which the gauge symmetry is already present in the 10-dimensional theory and upon compactification it may lead to chiral string models.
Local models are essentially the type II string models in which the gauge and matter fields are localised on D-branes. The extreme case could be the Standard Model localised on a D3 brane which is just a point in the extra dimensions. This allows to separate the physics questions between the questions that can be addressed only by how the Standard Model fits inside the D-brane and the global questions that do depend on the full structure of the six or seven extra dimensions. This is known in the string literature as the `bottom-up' approach to string model building.

\begin{center}
\begin{tabular}{ll}
\hline\noalign{\smallskip}
{\it Local Questions} & {\it Global Questions}\\
\hline\noalign{\smallskip}
Gauge Group & Moduli Stabilisation\\
Chiral Spectrum & Cosmological Constant\\
Yukawa Couplings & Supersymmetry Breaking\\
Gauge Couplings & Physical scales (unification, SUSY breaking, axions)\\
Proton Stability & Inflation or alternative, Reheating\\
Flavour issues (CKM. PMNS) & Cosmological Moduli Problem\\
\noalign{\smallskip}
\hline
\end{tabular}
\end{center}

The bottom-up approach is simply a systematic way to organise the challenge of realistic string model building. It is midway between traditional field theoretical model building and fully-fledged string constructions. Since the challenges are so big, it makes the search for realistic models more manageable asking a set of questions at the time and follow a modular approach to model building\footnote{At this stage this is only a convenient computational strategy rather than physically motivated. That will come later in this article.}. In contrast in global models such as the heterotic string all the physics questions have to be addressed at once. Heterotic models have other advantages such as starting already with  a unified group like $E_6$. 
Local models at the end have to be fully embedded into a complete compactification that is not straightforward. Both approaches have been followed with different amounts of success.

Over the years the main obstacles for realistic string model building have been, more than getting precisely the Standard Model spectrum at low energies, the stabilisation of moduli (which otherwise would source unobserved long-range interactions) and supersymmetry breaking. These are the questions that we will address next.

\section{Moduli Stabilisation}
This model independent sector for string compactifications is usually composed of the following fields: The axio-dilaton $S$, the complex structure moduli or size of the non-trivial 3-cycles $U_a$ and the K\"ahler moduli measuring the size of the non-trivial 4 and 2 cycles: $T_i$. Being scalar fields, their vevs have to be specified dynamically. The simplest supersymmetric compactifications leave them unspecified. The effective field theory depending on the K\"ahler potential $K$, superpotential $W$ and gauge kinetic function $f$ are such that the corresponding scalar potential is flat and here the SUSY non-renormalisation theorems protect the flatness of their potential perturbatively. Therefore there are limited sources of their scalar potential: 
\begin{enumerate}
\item{} Fluxes of Antisymmetric tensors to generate a non-vanishing tree-level superpotential.
\item{} Non-perturbative corrections to the superpotential $W$.
\item{} Perturbative (and non-perturbative) corrections to the  K\"ahler potential $K$ in both $\alpha'$ and string-loop expansions.
\item{} Induced $D$-terms.
\end{enumerate}

We will use all of them\footnote{Notice that the simplest constructions in which all these effects are neglected are not viable since neglecting all these effects is either inconsistent or very non-generic (setting by hand all fluxes to zero for instance). It is good news that including all available sources lead to more realistic physics.}. Fluxes are usually  complicated to deal with and it took many years before people learned to manage them \cite{gkp,dk}. The main reason is that they tend to change the structure of the compact manifold in such a way that there is no much mathematical understanding on the compactification manifold. Fortunately the case of IIB compactifications is such that the manifold after flux compactifications is conformal to the well studied  Calabi-Yau manifolds and this is one of the main reasons these compactifications have attracted much attention in the past decade. This is just a computational rather than conceptual advantage and from string dualities we know all other string compactifications should lead to similar physics once the technical aspects are sorted out. Since the spectrum of IIB supergravity has two 3-index antisymmetric tensors, fluxes on 3-cycles are able to fix all the $U_a$ fields and the axio-dilaton through a flux superpotential $W_0(S,U)$. Naively the fluxes of a three form field strength $H_3$ tend to fix the size $U$ of a three-cycle $\gamma$   by the quantisation condition $\int_\gamma H_3= 2\pi n$, this effect is captured in the EFT by the flux superpotential $W_0 (U,S)$\footnote{More explicitly the flux superpotential takes the form $\int G_3\wedge \Omega$ where $G_3=H_3+iSF_3$ with $H_3,F_3$ the two 3-form field strengths of the two  stringy 2-form potentials. Here $\Omega$ is the unique $(3,0)$ form that exists for every CY manifold. Expanding $\Omega$ in a basis of three-forms generates a superpotential dependence  on the $U_a$ fields.}.The perturbative superpotential cannot depend on the $T$ fields since their imaginary components are axion-like fields having a perturbative Peccei-Quinn shift symmetry: $\rm{Im} T_i\rightarrow \rm{Im} T_i + c_i$ and the holomorphicity of $W$ would then not allow dependence on the full superfield $T_i$. Therefore they can only appear in $W$ through non-perturbative effects.
\begin{equation}
W_{np}=\sum_i A_i e^{-a_iT_i}
\end{equation}
in which the $A_i$ may be functions of other moduli or even matter fields.

Combining this with the flux superpotential gives the full $W=W_0+W_{np}$ which combined with the corrections to $K$ are able to fix all moduli. This has been done in practice for only a handful
of models. 

The scalar potential derived from the general $\mathcal{N}=1$ supergravity expression $V=V_F+V_D$, with:
\begin{equation}
V_F=e^K\left[K^{I\bar{J}} D_IW D_{\bar{J}}\bar{W} - 3|W|^2\right]
\end{equation}
where $K^{I\bar{J}}$ is the inverse of the  K\"ahler metric $K_{I\bar{J}}=\partial_I\partial_{\bar{J}} K$ and $D_IW= \partial_I W + W \partial_IK$ is the K\"ahler covariant derivative. The $D$-term part of the salar potential is: 
\begin{equation}
V_D=\frac{1}{{\rm{Re}} f}\left( \xi_{FI}(T)+K_{\Phi}\mathcal{T} \Phi \right)^2
\end{equation}
where $\xi_{FI}\sim \partial K/\partial T$ are the (misnamed) field-dependent Fayet-Iliopoulos terms, only present for abelian groups, $\Phi$ a matter field transforming under the corresponding gauge group and $\mathcal{T}$ are the corresponding generators (charges in the case of a $U(1)$). Gauge indices suppressed.

Concentrating on the moduli dependence, the typical shape of the moduli scalar potential takes the form:
\begin{equation}
V_F\propto \left( \frac{K^{S\bar{S}} |D_SW|^2+ K^{a\bar{b}} D_aW \bar{D}_{\bar{b}}\bar{W}}{\mathcal{V}^2}\right)+\left(\frac{A e^{-2a \tau}}{\mathcal{V}}-\frac{Be^{-a\tau}W_0}{\mathcal{V}^2}+\frac{C |W_0|^2}{\mathcal{V}^3}\right)
\end{equation}

Here $\tau=\rm{Re}T$ represents a typical $T$ modulus, with $\mathcal{V}$ the overall volume (function of the $T$ fields) and the potential is meant to be seen as an expansion in large volume, where the effective field theory treatment is justified. In this case the first terms in parentheses are of order $1/\mathcal{V}^2$ and being positive definite they have to vanish at the minimum, imposing $D_SW=D_aW=0$ and therefore fixing $S$ and $U_a$ generically. This in turn fix the values of $W_0$ at the minima which is a huge distribution of values but mostly fitting in the range $0.1\leq |W_0|\leq 100$. The second parentheses is not positive definite and depending on the signs of the coefficients $A,B,C$ it gives a minimum for the K\"ahler moduli $T$. In particular the sign of $C$ depends on the sign of the Euler number of the Calabi-Yau manifold, by mirror symmetry half of them have negative Euler number and then positive $C$ implying a minimum at volumes of order 
\begin{equation}
\mathcal{V}\sim e^{a\tau}\qquad \rm{with} \qquad \tau\sim {\rm{Re}} ~S\sim 1/g_s> 1.
\end{equation}

 Implying an exponentially large volume. This gives rise to the LARGE volume scenario or LVS \cite{lvs}.
For very particular values of $W_0$, the  large number of solutions allows for a few of them to satisfy $|W_0|\ll 1$. In this case $W_0$ can be tuned so that $W_0\sim W_{np}\sim e^{-a\tau}$ so the term proportional to $C$ can be neglected and a minimum can be found for $\tau$. This is essentially the KKLT scenario \cite{kklt}. But for generic values of $W_0$ only the LVS works. It has been shown that this holds as long as the number of 3-cycles is larger than the number of 4-cycles and both greater than one ($h_{12}>h_{11}>1$) which is satisfied for half of the CY manifolds by mirror symmetry. The second condition is the existence of at least one collapsible 4-cycle which is the generic case.

In both KKLT and LVS the position of the minimum is at negative values of $V_F$ so leading to AdS vacua. The main difference is that in KKLT this minimum is supersymmetric ($D_TW=0$) but in LVS supersymmetry is broken. They both have small parameters in which to base the approximate effective theory, $1/\mathcal{V}$ for LVS and $W_0$ for KKLT. Notice that we have not yet used the $D$-term part of the scalar potential. This is more model dependent since it depends on charged matter fields for which we need to specify the concrete model. Being positive definite it will tend to uplift the minimum  found from purely $V_F$ \cite{bkq}. However for KKLT there is a strong restriction, since the minimum is supersymmetric, it means all $F$-terms vanish which in turn imply that the $D$-terms have to vanish. In LVS $D$-terms can lift the minimum opening the possibility of leading to de Sitter space. But  this is very model dependent and there is a need to have a full global compactification with all matter fields to make a proper study. This will be addressed later on in this article.

Another way to uplift the minimum to de Sitter was proposed in KKLT by introducing anti D3 branes at the tip of a warped region in the compact manifold. This provides a positive contribution to the vacuum energy given by the warped brane tension and an explicit supersymmetry breaking source. The effective field theory is more complicated to handle since it leaves the regime of validity of $\mathcal{N}=1$ supergravity. For LVS this is also an option but $D$-terms (and matter $F$-terms) provide a more promising avenue to obtain de Sitter within a purely supersymmetric effective action \footnote{For other interesting proposals to obtain de Sitter from purely supersymmetric EFTs see \cite{desitter}.}. 

In LVS there is a clear hierarchy of scales shown in the table below.

\begin{center}
\begin{tabular}{ |l|l|} 
\multicolumn{2}{c}{\it Relevant Physical Scales in LVS} \\
\hline
Physical scale & Volume dependence\\
\hline
Planck mass & $M_P$\\ 
String scale & $M_s=\frac{M_P}{\mathcal{V}^{1/2}}$\\ 
Kaluza-Klein scale & $M_{KK} = \frac{M_P}{\mathcal{V}^{2/3}}$\\ 
Gravitino mass & $m_{3/2} = \frac{M_P W_0}{\mathcal{V}}$\\ 
Volume modulus mass & $m_V= \frac{M_P W_0}{\mathcal{V}^{3/2}}$\\ 
`Fibre' moduli mass & $m_F = \frac{M_P W_0}{\mathcal{V}^{5/3}}$\\
\hline
\end{tabular}
\end{center}

Notice that a clear bound for $W_0$ is $|W_0|\ll \mathcal{V}^{1/3}$ in order to have a proper hierarchy ($M_{KK}\gg m_{3/2}$) and guarantee the consistent use of an effective field theory to describe the physical implications of the scenario (an even stronger bound is $|W_0|^{1/6}$ guarantees the effective potential being smaller than $M_{KK}^4$) \cite{lvs,wbounds}. Also even though the gravitino mass is supposed to set the scale of all particles that receive a mass after SUSY breaking, all moduli $S,U_a$ and most of $T_i$ receive a mass of the order of the gravitino mass, however the overall volume modulus has a mass much lighter and remains small after quantum effects even though it is not protected by supersymmetry
\cite{cmq}. Furthermore, in some Calabi-Yau manifolds which happen to be fibrations of 4D manifolds such as K3, the corresponding modulus does not receive a mass until loop effects are taken into account and therefore their mass is even smaller than that of the volume modulus ($m\sim W_0/\mathcal{V}^{5/3}$).

Some general properties of LVS:
\begin{itemize}
\item{\it Stability.} Even though the overall minimum is locally stable the fact that even the AdS vacuum is not supersymmetric makes it subject to non-perturbative instabilities, such as bubble of nothing decay. This was studied in \cite{stability}. As long as the effective field theory is valid the AdS minimum is stable and no indication to a bubble of nothing decay. This leads to the possibility of having a CFT dual and therefore a proper non-perturbative description of these vacua despite being non-supersymmetric. The dS minima are clearly metastable and the decay rate goes like $\Gamma\sim e^{-{\mathcal{V}}^3}$. The probability to decay to an AdS minimum is preferred over a dS as a ratio
$P_{dS}/P_{adS}\sim e^{-{\mathcal{V}}}$ whereas its decay towards the 10D decompactification vacuum ($\mathcal{V}\rightarrow \infty$) is further suppressed $P_{dec}/P_{dS}\sim e^{-{\mathcal{V}}^2}$. Clearly the larger the volume the more stable the vacuum.

\item{\it Bounds on the volume.} However the volume cannot be arbitrarily large since for values $\mathcal{V}\sim 10^{30}$ the string scale becomes smaller than the TeV scale, also beyond $\mathcal{V}\sim 10^{15}$
the gravitino mass (and usually soft terms) will be smaller than the TeV scale. Finally for volumes $\mathcal{V}\geq 10^{9}$ the volume modulus becomes lighter than $10$ TeV which would lead to the cosmological moduli problem (CMP). Smaller volumes $10^3<\mathcal{V}\geq 10^8$ are consistent and survive overclosing (with the larger volumes being the more stable from the previous item) but still imply a special cosmological role for the volume modulus (or any lighter one in particular cases). This modulus is the latest to decay and its decay would be the source of reheating of the observable universe leading to interesting post-inflationary cosmology (see for instance \cite{darkradiation}).

\item{\it Inflation.} The three terms in the second parentheses for $V_F$ hint at a concrete realisation of inflation. Assuming the volume is already at its minimum value, the potential for $\tau$ is precisely of the form $A-Be^{-x}$ for large values of $\tau$ which is one of the preferred inflationary potentials for a canonically normalised inflaton field $x$. In order to achieve this concretely at least three $T_i$ fields are required which is very generic in string compactifications. Loop corrections may destabilise the flatness of the potential during inflation. A more elaborated and stable under quantum corrections model of inflation has been proposed in which the inflaton is a fibre modulus.
For this scenario the spectral index and tensor to scalar ratio $r\sim 10^{-3}$ falls just in the preferred Planck regime (see \cite{planck} for a recent overview). However, if the recent results from BICEP are confirmed $ r\sim \mathcal{O} (0.1) $ then these scenarios are ruled out by experiment. An example on how string scenarios can be predictive and contrasted with experiment. 
The string scenarios consistent with BICEP: N-flation, axion monodromy  and Wilson line inflation \cite{larger} can be embedded in the LVS. More work in this direction is needed.

\item{\it Axions.} There are plenty of axions in string compactifications, many can survive at low energies but some do not. In LVS it is clear that the axion partners of the dilaton and K\"ahler moduli stabilised by non-perturbative effects acquire a mass of order the gravitino mass. Other axions are eaten by anomalous $U(1)$s by the Stuckelberg mechanism. But some survive at very low energies, in particular the axion partner of the volume modulus is essentially massless after moduli stabilisation and may have some implications for late time cosmology. In particular contributong to dark radiation. Also axions coming from phases of matter fields may survive low energies but are more model dependent (see \cite{michele} for a recent overall review on stringy axions).

\end{itemize}

Before finishing this section let us also mention the progress made in other string compactifications.
In heterotic strings despite many efforts there is no yet a compelling scenario for stabilisation of all moduli. Substantial progress has been made in the past few years to stabilise most of them.
Contrary to IIB strings that have two 3-index antisymmetric tensors to turn on, in the heterotic there is only one and so fluxes are not as efficient as they are in IIB strings. The number of equations and unknowns is similar leaving no room for a landscape and no mechanism to solve the cosmological constant problem. Furthermore they move the model away from the Calabi-Yau spaces. Yet, since heterotic strings carry a large gauge group already before compactification, this introduces new moduli (called `bundle moduli') that can actually help to fix the complex structure moduli by consistency gauge conditions that have to be satisfied. Nonperturbative effects still can fix the K\"ahler moduli, similar to the IIB case. Clearly further progress is expected in this direction since for realistic model building heterotic models are probably the most developed.

G2-holonomy manifolds compactifications of the 11D supergravity limit of M-theory have been studied also. There is no explicit model of particle physics from these compactifications which need much more mathematical developments although they are also clearly local models also. For moduli stabilisation there are interesting properties that can be extracted without entering into details. In particular all moduli are similar to the $T$ moduli of type IIB, making the moduli stabilisation issue easier to define. Fluxes are not an option in this formalism and therefore the full superpotential is non-perturbative. A superpotential of the form
$W=\sum_{ij} A_i e^{-a_{ij}T_j}$ has been proposed with the potential to fix all $T$ fields. The general properties of this scenario have been summarised in \cite{akk}.

\section{Supersymmetry Breaking}

The breaking of supersymmetry is intimately related to moduli stabilisation. This explains that only after a well defined framework for moduli stabilisation it was possible to extract information about supersymmetry breaking in string theory. Progress in moduli stabilisation eventually extends to progress in supersymmetry breaking. Contrary to moduli stabilisation in which to large extent the location of the standard model can be ignored, here it is fundamental. In IIB local models the standard model can be inside a D3 brane at a singular point in the extra dimensions or at a D7 brane wrapping a 4-cycle of the extra dimensions. Other odd dimensional branes are dual to these and even dimensional branes would appear in the IIA case.

What we can see from both LVS and KKLT\footnote{For KKLT even though the source of SUSY breaking is the uplift by anti D3 branes a very interesting scenario called mirage mediation has emerged with interesting phenomenological implications \cite{mirage}.} is that both the $S$ and $U_a$ fields do not break supersymmetry at leading order since they are fixed by the condition $D_SW=D_aW=0$.

An important information is regarding the contribution to supersymmetry breaking of the cycle where the standard model lives on the D7 brane case. On the D3 brane case this may also be thought as a collapsed 4-cycle. The point is the following. In any brane sector where there is chiral matter, the corresponding $T$ modulus, measuring the size of the 4-cycle that the D7 brane is wrapping, acquires a charge under an anomalous $U(1)$. Therefore it is not possible to have a term of the form $W_{np}=Ae^{-aT_{SM}}$ in the superpotential in which $A$ is a constant, since this term would not be gauge invariant. Therefore whenever a dependence on $T$ appears it has to come together with a dependence of $A$ on charged matter fields that compensate the gauge variation of $T_{SM}$. This makes it very difficult to stabilise $T_{SM}$ by the LVS or KKLT methods since the SM fields are supposed to have zero vev at the high scales (otherwise they may induce  colour breaking for instance). Let us call this the BMP constraint \cite{bmp}. As long as one of the moduli $T_h$ from a hidden sector (in which chirality is not required) appears in $W_{np}=Ae^{-aT_h}$ with constant (or only moduli dependent) $A$ the LVS minimum is obtained. The $T_{SM}$ cycle may be fixed by D-terms or even by loop corrections to $V$.

Some general properties of SUSY breaking in LVS are

\begin{enumerate}
\item{} The source of SUSY breaking is well identified coming from generic values of the 3-form fluxes for which $D_SW=D_aW=0$ but $W_0\neq 0$ and the $F$- term of the K\"ahler moduli is non-vanishing. This is major difference as compared to KKLT in which before uplifting SUSY is not broken and its breaking is fully determined by the anti-D3 brane that performs the uplift. In LVS the uplift can be done in different ways but even if the anti-brane is added, its contribution to soft terms is usually negligible.

\item{} The existence of the landscape allows for the first time to address simultaneously the cosmological constant and the hierarchy problems. This justifies  the standard strong assumption that has been made over the years regarding the use of low-energy SUSY to address the hierarchy problem: that something else takes care of the cosmological constant problem and has no direct influence on the calculation of soft SUSY breaking terms (\emph{good}). But, by the nature of the landscape, it prevents us to find new physical phenomena at low energies determined by the cosmological constant (\emph{bad}). It also opens the possibility to use  anthropic arguments to address the hierarchy problem (\emph{ugly}).

\item{} SUSY is broken by the K\"ahler moduli which do not enter in the tree-level matter superpotential, therefore as long as K\"ahler and complex structure moduli do not mix (true at tree-level K\"ahler potential) then flavour problems, generic for gravity mediation, are ameliorated \cite{soft1,mirror}. The correct estimate on how much flavour violation is induced by quantum corrections mixing the moduli is an open question.

\item{} The dominant source of SUSY breaking in the EFT  is the F-term of the volume modulus. But this gives rise, to leading order, a no-scale model with vanishing soft terms. Therefore next order corrections are relevant. In order to explicitly compute the soft terms requires knowledge of the F-term of the cycle that hosts the SM. If its F-term vanishes (to avoid the BMP obstruction) then soft terms can be very much suppressed ($M_{1/2}\sim \mathcal{O} (m_{3/2}/\mathcal{V})$). Otherwise they are proportional to the gravitino mass up to a small loop factor. We list the different scenarios in the table below.
Notice that if we use a TeV gravitino mass, it would select one of the first two scenarios. The second one needs a strong fine-tuning in $W_0$ in order to simultaneously obtain the unification and SUSY breaking scales at the preferred values. The first one does not need the tuning at the cost of lowering the unification scale. Both suffer from the cosmological moduli problem (CMP). If we use the avoidance of this problem as the selection criterion then the last three scenarios are preferred. The first one of those gives up a natural explanation of the TeV scale. The last two are sequestered scenarios in which the $F$ term of the SM modulus vanishes to avoid the BMP obstruction. Sequestered scenarios may be subject to modifications due to quantum corrections and at the moment only models in which the SM is at D3 branes at singularities seem to remain truly sequestered. If so then both TeV SUSY breaking and the preferred GUT scale can be obtained without the CMP.
\end{enumerate}

\begin{center}
\begin{tabular}{|c|c|c|c|c|c|} 
\multicolumn{6}{c}{\it SUSY Breaking Scenarios  in LVS} \\
\hline
Scenario & String Scale &  $W_0$ & $m_{3/2}$ & Soft masses & CMP\\ \hline
Intermediate Scale & $10^{11}$ GeV  & $\mathcal{O}(1)$ & $1$ TeV & $M_{soft}\sim1$ TeV & Yes\\
Tuned GUT Scale & $10^{15}$ GeV &  $10^{-10}$ & $1$ TeV & $M_{soft}\sim 1$ TeV & Yes\\
Generic GUT Scale & $10^{15}$ GeV &  $\mathcal{O}(1)$ & $10^{10}$ GeV & $M_{soft}\sim 10^{10}$ GeV & No\\
Sequestered Unsplit & $10^{15}$ GeV &  $\mathcal{O}(1)$ & $10^{10}$ GeV & $M_{soft}\sim \frac{m_{3/2}}{\mathcal{V}}\sim 1$ TeV & No\\
Sequestered Split & $10^{15}$ GeV &  $\mathcal{O}(1)$ & $10^{10}$ GeV & $M_{1/2}\sim  \frac{m_0}{\mathcal{V}^{1/2}}\sim\frac{m_{3/2}}{\mathcal{V}}\sim 1$ TeV & No\\
\hline
\end{tabular}
\end{center}

The intermediate scale scenario and the tuned GUT scenario have been studied in the past with some detail \cite{soft1}. Despite the fact that they both have the CMP the soft terms can be calculated in a more explicit way since the dominant contribution comes from the $F$-term of the SM cycle. The generic GUT scale scenario has not been studied in detail since the superpartners are much heavier than the TeV scale and no hope to be detected not even in the long term. This scenario however realises explicitly the large SUSY scale proposals recently discussed \cite{ibanez,hall} in which the hierarchy problem is not solved by low-energy supersymmetry but fits with the measured value of the Higgs mass. The two sequestered scenarios proposed in \cite{soft2} are very interesting because they both have the preferred GUT scale while at the same time superpartners have the TeV scale that solve the hierarchy problem. However since the SM cycle does not break SUSY then the explicit expressions for the soft terms are difficult to compute since they are small compared to the gravitino mass and they are more model dependent. In particular they depend on the uplifting mechanism which gives negligible contributions in the other scenarios. Therefore these scenarios have not been studied in detail. See however \cite{aparicio}.

\section{Local and Global Model Building}

One of the implications of the LVS is that the standard model has to be localised. The reason is that if it lives on a D7 brane wrapping a four-cycle, this cycle cannot be the one dominating the volume, since the volume is exponentially large and the gauge coupling of the gauge theory living on the brane is inversely proportional to the size of the cycle $g_{SM}^{-2}\sim \rm{Re} T_{SM}$ and would generically be too small to fit realistic values $\mathcal{O}(20)$ expected at the GUT scale. Therefore either the SM lives on a D7 brane wrapping a small cycle or at a D3 at a singularity. In both cases it is localised. This provides an independent  argument to consider local string models in IIB compactifications. Notice that local F-theory models can be seen as strong-coupling generalisations of magnetised D7 brane models and in principle also fit with this analysis. 

We will restrict here to  local models at singularities (for a more comprehensive discussion see the nice review \cite{mp} in which both local F-theory models and branes at singularities are reviewed). An argument to justify this selection is the following. Since the SM is chiral the corresponding modulus $T_{SM}$ is usually charged under anomalous $U(1)$  groups on the brane. This has two important implications, first as mentioned before there is the BMP obstruction to fix $T_{SM}$ from nonperturbative effects. Second the corresponding $U(1)$ group has a Fayet-Iliopoulos (FI) term $\xi \propto {\rm Re} T$.  The corresponding D-term potential 
\begin{equation}
V_D\propto \left(\xi -q_i|\phi_i|^2\right)^2
\end{equation}
 combined with soft mass terms induced for matter fields $m_i^2|\phi_i|^2$ tend to prefer $\xi=0$ and then towards a collapsed cycle ${\rm Re} T_{SM}\rightarrow 0$. This implies that the effective field theory (EFT) valid for large values of moduli (compared with the string scale) is not valid and a different EFT has to be used valid as an expansion around the singularity. Fortunately this is also known for orbifold-like singularities in which the FI term is also proportional to the size of the blow-up mode $\xi\propto \rho_{SM}$ and again the D-term minimisation tends to prefer the collapsed cycle $\rho_{SM}\rightarrow 0$ \footnote{Notice that EFTs can be written in the two regimes in which the cycle is either much larger than the string scale or very close to zero. It is a complicated open question to find the matching between the two EFTs going through the domain of string scale cycle.}. Quantum corrections may blow-up the singularity to non-vanishing values of the blow-up mode but in general keeping it within the singularity regime ($i.e.$ $\rho_{SM}\leq l_s^4$) with $l_s$ the string length scale.

This argument is however not a proof that the SM has to live at a singularity since there are two ways out: the FI term is model dependent and usually is a linear combination of moduli fields. There may be a way to engineer the models so that these combinations vanish with non-vanishing fields (see for instance \cite{cmv}). Furthermore the soft terms contributions to the matter fields $\phi_i$ may be tachyonic and then $\phi_i\neq 0$ at the minimum. 

Local models of D-branes at singularities have been studied over the years \cite{aiqu,verlinde,sven}. But it is only until very recently that they have been systematically embedded in compact Calabi-Yau compactifications including moduli stabilisation \cite{global} (see also \cite{per}).

Let us start with the fully local constructions first and discuss the global embedding later.
The gauge theory of branes at singularities can be described by quiver diagrams with nodes and arrows, node $i$ represents $n_i$ D-branes implying a group $U(n_i)$ and arrows going from the $i$ node to the $j$ node represents a bifundamental $({n}_i, \bar{n}_j)$. Usually a closed loop in a quiver represents a gauge invariant superpotential term although the precise structure of the superpotential needs further techniques based on dimer diagrams that we will not discuss here \cite{sven}. A simple example is provided by the $\mathbb{Z}_3$ singularity with a triangular quiver and three arrows connecting the nodes. Choosing $n_i=n_j$ guarantees an anomaly free model (except for an anomalous $U(1)$ that become massive from the Stuckelberg mechanism) which can be easily evaluated by counting the number of arrows coming in and out each node which should match for anomalies to be canceled. We show in the figure the simplest of these cases including the SM which corresponds to $n_i=3$ and is precisely the trinification model $SU(3)^3$ with three families.
\begin{equation}
3\left[(3,\bar{3},1)+ (1,3,\bar{3})+ (\bar{3}, 1, 3)\right]
\end{equation}
which is actually three families of $27$s of $E_6$.

\begin{figure}
\centerline{\includegraphics[height=5cm]{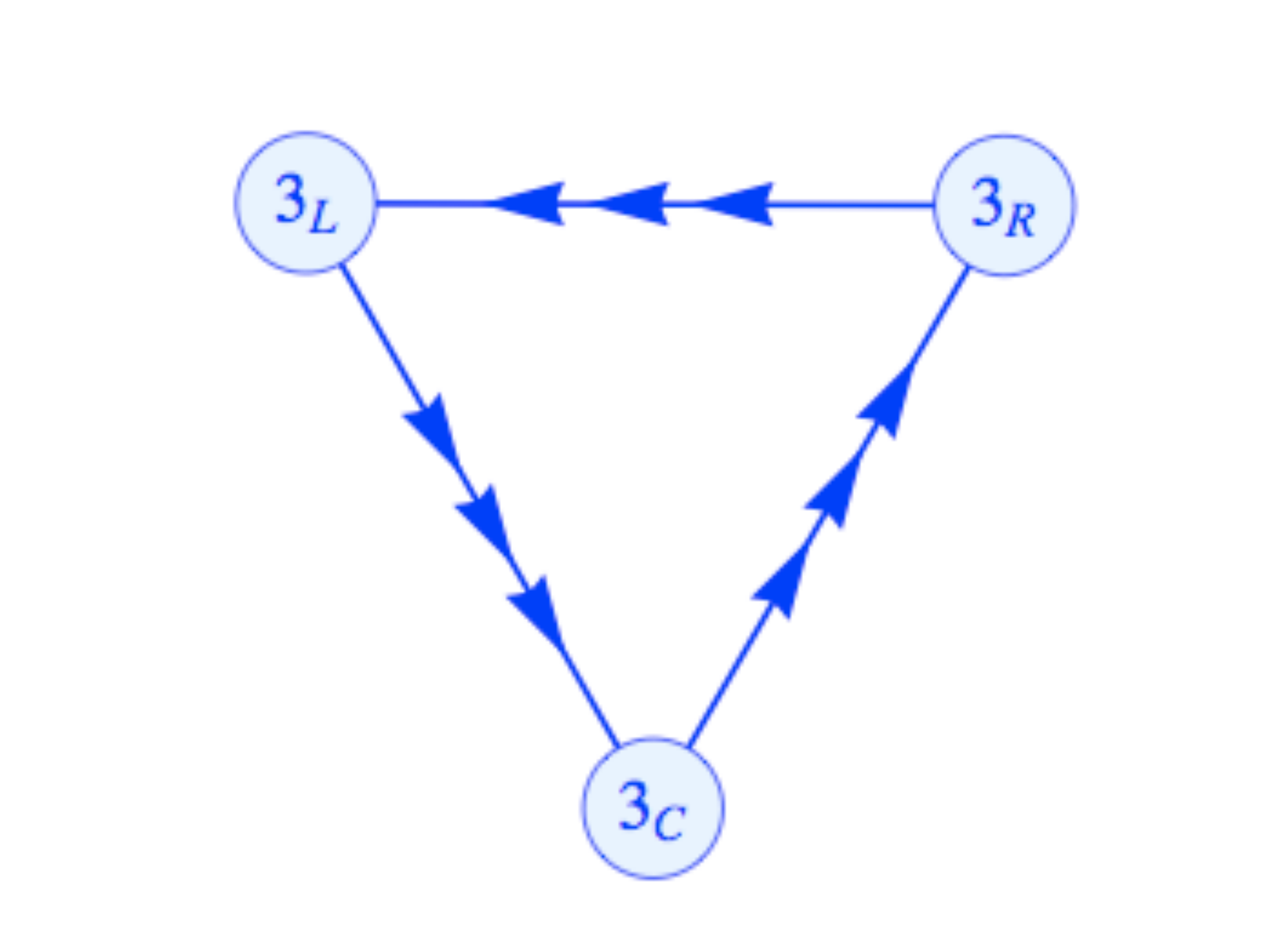}}
\caption{The simplest realisation of a chiral model at D3 brane singularities containing the SM (the trinification model at $dP_0$ in this case). The three gauge groups are identical, $SU(3)^3$. The three arrows imply three families of bi-fundamental fields. The equal number of ingoing and outgoing arrows at each node guarantees anomaly cancellation.}
\label{ra_fig2}
\end{figure}

This simple orbifold singularity is actually the simplest of a very special class of singularities called del Pezzo $n$ or $dP_n$ singularities. These singularities are actually collapsed del Pezzo surfaces which are four-dimensional surfaces (which in our case are four-cycles inside a Calabi-Yau space). These are defined as the complex 2-dimensional projective space $\mathbb{P}_2$ blow up at $n$ points with $n=0,1\cdots 8$. The $\mathbb{Z}_3$ singularity corresponds to $dP_0$. The special property of the del Pezzo surfaces is that they are the simplest 4-cycles that can collapse to a single point in a Calabi-Yau manifold. The quiver diagrams and superpotential couplings have been uncovered on the past years. Leading to phenomenological considerations. The corresponding quiver diagrams have $n+3$ nodes. For instance the next case, the $dP_1$ singularity has a square quiver. Furthermore given a vev to a bifundamental between two quivers merge the two  nodes into one and reproduces the $dP_0$ case. This can be interpreted geometrically as the fact that $dP_n$ has $n+1$ non-trivial 2-cycles and the higgsing corresponds to an independent collapse of one of these cycles.

Other more general classes of singularities have been studied (called toric singularities) that provide a large number of local string models with chiral matter content.

Some of the general properties of these local models are the following:

\begin{itemize}

\item{} The gauge coupling at the singularity is given at tree level by the vev of the dilaton field $S$ so there is unification without necessarily having a simple GUT group. The gauge group is usually a product of simple groups. This avoids the standard problem of D-brane models for which having simple GUT groups leads to vanishing top Yukawas (which has been the main argument to consider F-theory models). Top Yukawas are easily generated if the groups are not simple.

\item{} The maximum number of families (arrows) happens to be three, which is the one observed in nature. However starting with a complicated enough quiver and by higgsing  more (but not many) families may be obtained.

\item{} There is always one zero eigenvalue in the spectrum. In the case of $dP_0$ the massive states have eigenvalues $(M,M,0)$ where the mass $M$ is determined by the vev of the low-energy Higgs field. Having two families degenerated and one massless is not realistic. However all other $dP_n$s with three families have eigenvalues $(M,m,0)$  with $m\ll M$ which have the proper hierarchy observed in nature. Quantum effects are expected to lift the zero-eigenvalue although this is not straightforward \cite{bkmq}.

\item{} Apart from the overall $U(1)$ that can be simply decoupled, there are two anomalous $U(1)$s that obtain their mass by the Stuckelberg mechanism.

\end{itemize}

The rest of the physics has to be extracted in a model by model case. A large increase on the number of these models comes from  the possibility to add D7 branes to each configuration. These allow for many more choices of integers $n_i$ (for instance in $dP_0$ we can have $(n_1,n_2,n_3)=(1,2,3)$ giving the SM gauge group from D3 branes times a hidden or flavour symmetry coming from the D7 branes which cancel the anomalies). The number of D7 branes is restricted by tadpole cancellations which usually are equivalent with anomaly cancellation. This enhance substantially the number of realistic models at singularities.

Quasi realistic models including the SM gauge and matter fields as well as proper Yukawas, CKM  and PMNS mixings, proton stability, and unification have been constructed using D3/D7s at singularities. In particular $dP_3$ models have proved to be manageable enough and at the same time rich enough to address the flavour questions \cite{sven}.
The issue of gauge coupling unification in these models is no much if there is unification (up to small thresholds as usual) which is automatically achieved  but knowing that the gauge couplings are unified how do they evolve to low energies. Since usually the matter content of the models is not just that of the MSSM then the success of the MSSM for gauge unification would be most probably an accident in this class of models (for good or for bad as in most  string constructions). Achieving the right values of the couplings at low energies is not easy. There is one simple local model that stands out in this regard. This is the one based on $(n_1,n_2,n_3)=(3,2,2,)$ giving rise to a left-right symmetric model $SU(3)\times SU(2)_L\times SU(2)_R\times U(1)_{B-L}$ with three families and  three extra pairs of Higgses (after D7-D7 states get vevs):

\begin{equation}
3[(3,\bar{2},1) + (\bar{3},1,2) + (1,{2},1)+ (1,1,\bar{2})+ (1,2,\bar{2})] + \rm{singlets}
\end{equation}

The extra matter content together with the fact that the hypercharge normalisation is not standard combine in a way that there is unification with a similar level of precision as the MSSM. The unification scale though is intermediate $10^{11}$ GeV approximately
\cite{aiq, aiqu}. In order to get realistic Yukawas this model can be embedded into higher $dP_n$s.


This model built from both D3s and D7s and the trinification model built from only D3s are examples of very simple quasi-realistic chiral models. But being purely local they cannot be called string vacua, since in their construction compactification was not considered. In order to become honest-to-God string compactifications the corresponding singularity has to be embedded in a compact Calabi-Yau manifold (to preserve supersymmetry and chirality). Actually starting from IIB string theory and compactifying on a CY manifold leads to $\mathcal{N}=2$ supersymmetry and then non-chiral models. The missing ingredient is {\it orientifolding}. This is essentially a $\mathbb{Z}_2$ twist of the CY compactification exploiting the fact that the worldsheet theory is orientable and has a $\mathbb{Z}_2$ symmetry. It is well known that combining this twist with a CY compactification leads to chiral $\mathcal{N}=1$ theory in 4D. 

A concrete way to embed local singularity models in fully-fledged CY compactifications was outlined in \cite{cmq}. The idea is to look for CY compactifications with at least three $dP_n$ surfaces. Two of them map to each other under the $\mathbb{Z}_2$ orientifold twist, where the SM would live and a third one to provide the non-perturbative correction to the superpotential. A fourth 4-cycle (not a $dP_n$) will be the one dominating the volume. This is the minimum set-up. In the figure below we illustrate it. 

\begin{figure}
\centerline{\includegraphics[height=7cm]{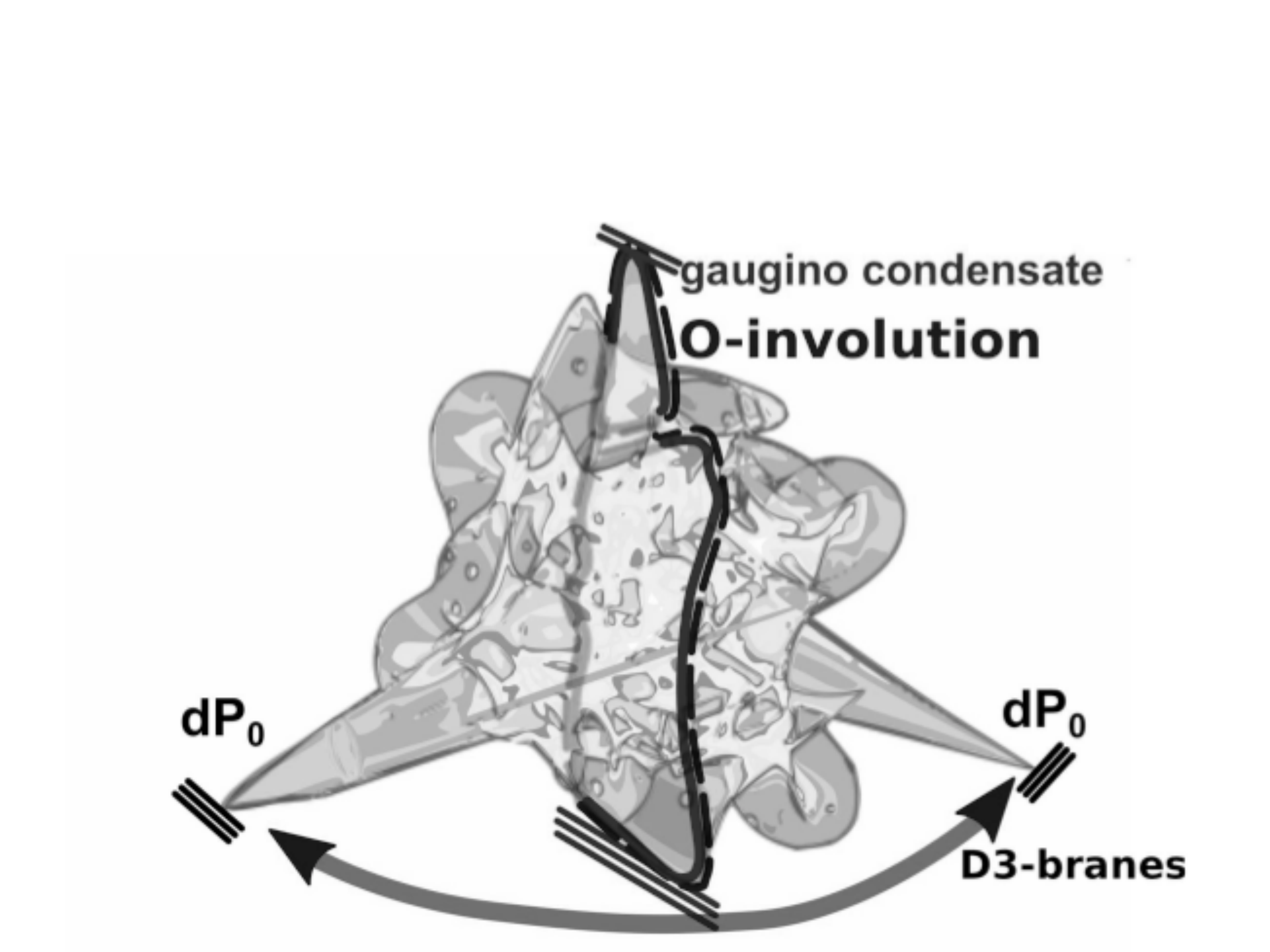}}
\caption{An explicit global embedding of local D-brane models. The SM is located at a $dP_0$ mapped to an identical singularity by a $\mathbb{Z}_2$ twist. Two ther 4-cycles are needed to stabilise K\"ahler moduli and obtain a global realisation of LVS with chiral matter.}
\label{ra_fig5}
\end{figure}

In the past 2 years concrete realisations of these models were achieved \cite{global}. The details are too technical for this review but it is worth emphasising the main points:
\begin{enumerate}
\item{} A classification of Calabi-Yau manifolds constructed as hypersurfaces from toric varieties is available from the work of Kreuzer and Skarke \cite{ks}. From this large class of models a classification of those with a relatively small number of K\"ahler moduli (4 and 5) to fulfill the requirements. This gives a few thousand models of which a couple of hundred have $dP_n$ surfaces mapped into each other under a $\mathbb{Z}_2$ (illustrated in the figure)
\item{} A configuration of D3 and D7 branes as well as orientifold planes (fixed under orientifold action) is introduced satisfying a highly non-trivial set of consistency conditions: tadpole cancellations (local and global) for all D-brane charges, cancellation of the so-called Freed-Witten anomalies that appear in the presence of fluxes, K-theory charges, etc.
\item{} The set-up allows for the SM to be hosted at the singularities on the $dP_n$'s mapped into each other. Realising globally the local examples mentioned before.
\item{} Furthermore, the conditions for moduli stabilisation are realised with non-perturbative superpotential generated at the third $dP_n$ cycle (either by Euclidean D3 branes or gaugino condensation. 
\begin{figure}
\centerline{\includegraphics[height=4cm]{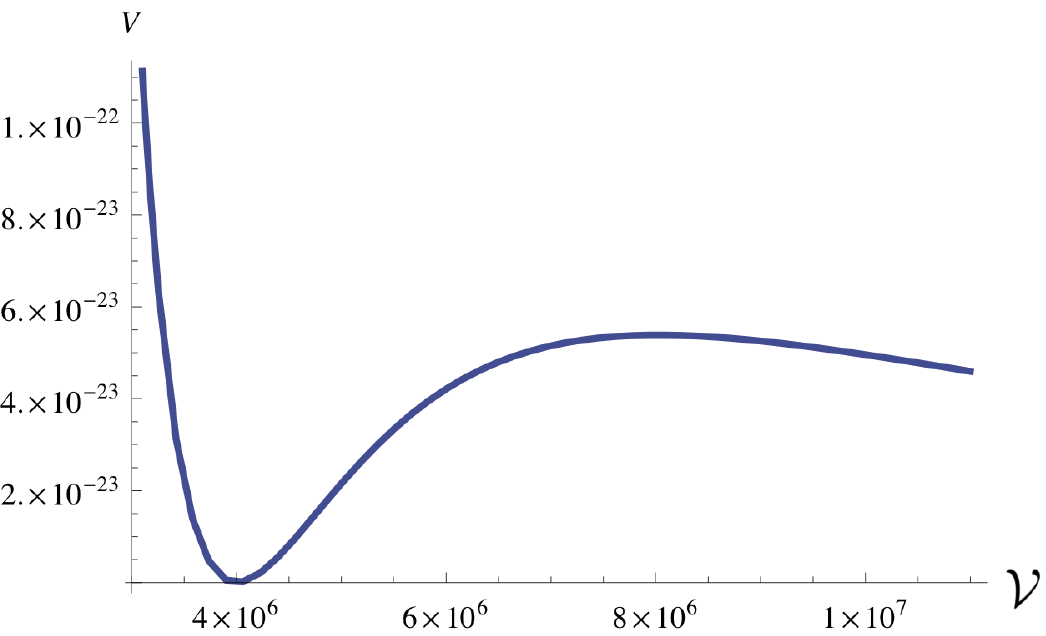}}
\caption{de Sitter minimum for a compact fluxed CY compactification with chiral matter at a $dP_0$ singularity.}
\label{ra_fig5}
\end{figure}
\item{} The modulus corresponding to the SM cycle is stabilised by D-terms in a way that it is naturally stabilised at zero and therefore in the singularity regime. This part of the potential takes the form $V(T_{SM},\phi)=V_D+ V_{soft}$ with $V_D\sim (\xi_{FI}+q|\phi|^2)^2$ and $V_{soft}\sim m^2|\phi|^2$. It is clear that as long as the soft masses for matter fields $\phi$ are non-tachyonic this fixes the minimum (at leading order ) at $\phi=0$ and $\xi_{FI}=0$ since both terms in the potential are positive definite. This is only leading order since these terms are of order $1/\mathcal{V}^2$ so small vevs are generated that compete with  next order in $1/\mathcal{V}$ expansion  which is the $1/\mathcal{V}^3$ that precisely gives rise to the LVS. These, together with the F-term of the matter fields $\phi$ end up contributing a term of order $\delta V= c W_0^2/\mathcal{V}^\alpha$ with $1\leq \alpha< 3$ and $c>0$. This term uplifts the LVS minimum to higher values including de Sitter space and the superpotential can be tuned to give an almost Minkowski vacuum. Notice that this tuning of $W_0$ is easy to achieve knowing that the large number of complex structure moduli  allows for order $10^{500}$  solutions for values of $W_0$ in the small range $0.1<W_0<10^3$.
\item{} Having such explicit CY compactifications allows also to explicitly solve the complex structure and dilaton equations $D_UW=D_sW=0$. However technically this is a huge number of solutions and it has not been done for fluxes in all the 3-cycles. Nevertheless some CYs admit enough discrete symmetries that allow to fix most of the complex structure moduli and only using fluxes in a handful of them. This has been done recently in \cite{complexstructure,complexkahler}. Then we have for the first time a global CY compactification \cite{complexkahler} with quasi-realistic visible sector including the SM, with all geometric moduli stabilised leading to de Sitter space using a fully supersymmetric EFT. Supersymmetry broken with computable soft terms.
\end{enumerate}
As a non-trivial check of `phenomenological consistency' the local LR symmetric model sketched before for a $dP_0$ singularity has an interesting phenomenological property: having the LR spectrum plus three pairs of Higgses, together with the normalisation of the hypercharge gives precise unification at an intermediate scale $10^{11}$ GeV with an accuracy similar to that one for the MSSM. However from the global realisation, both the unification scale and the value of the coupling at unification should be outputs of the dynamics of the global model after moduli stabilisation. It is in fact remarkable that for the global realisation of this model the volume is found to be of order $\mathcal{V}\sim 10^{12}$ and the gauge coupling (dilaton vev) precisely of the value that fits with the unification. Two completely independent calculations give rise the same physical quantities (see the second reference in \cite{global} for details). Rather than emphasising the qualities of this model (that has other phenomenological problems regarding flavour and the CMP) this illustrates the challenge for any other attempt to achieve unification including moduli stabilisation.

\begin{figure}
\centerline{\includegraphics[height=5cm]{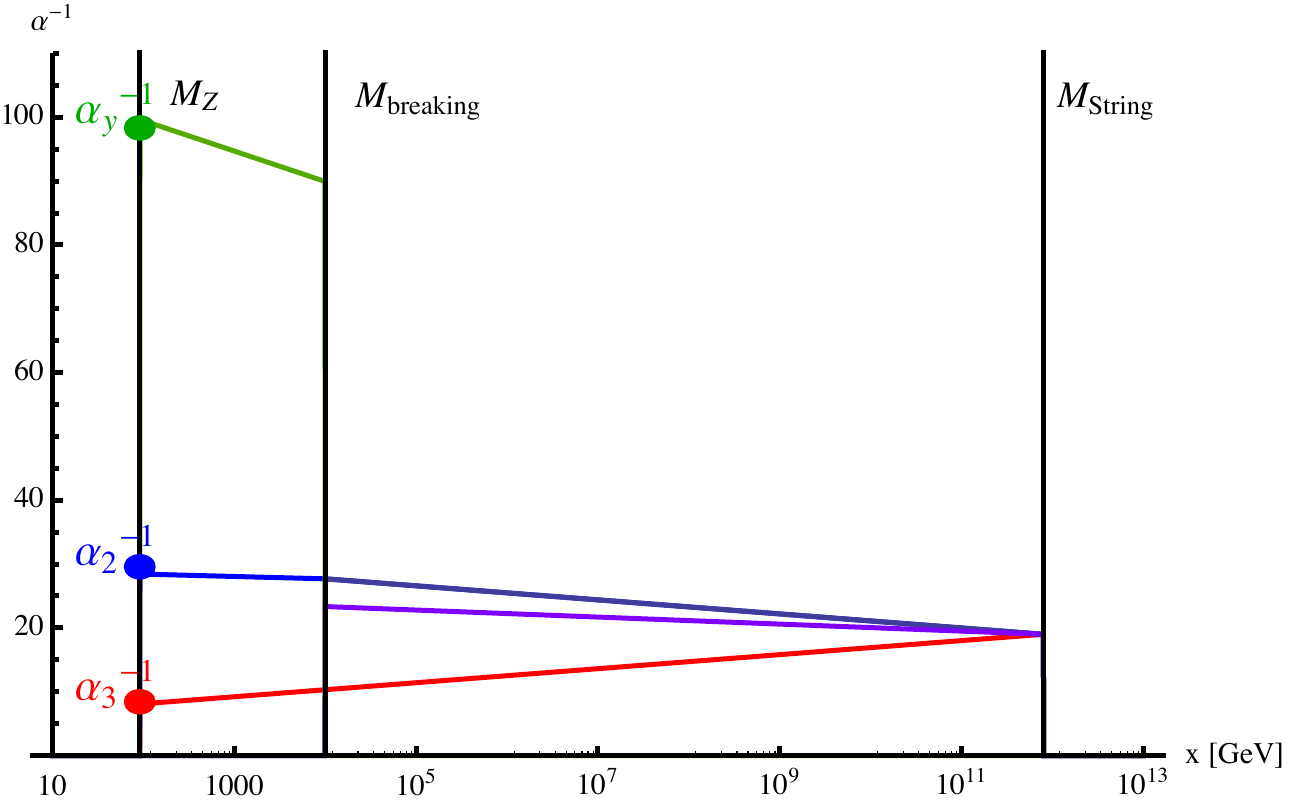}}
\caption{Unification in Left-Right symmetric model with low energy RG running matching the unification scale and gauge coupling with those values obtained independently from moduli stabilisation.}
\label{ra_fig1}
\end{figure}

\section{Open Questions}

Clearly local models have passed a threshold in the sense that there are now explicit local chiral models that have been properly embedded in a global compactification including moduli stabilisation and realistic values of physical scales, including de Sitter space from a fully supersymmetric effective field theory  and computable soft SUSY breaking terms. This is already impressive progress and a  
 sign of a healthy field that has evolved over the years with continuous progress. Still obtaining a fully realistic model is an open question.

On more concrete open questions:
 Embed the most realistic local models in a global compactification, such as those based on $dP_3$ or even higher $dP_n$ including a successful inflation scenario.  Also in cosmology, address in detail post-inflationary cosmological issues such as reheating, dark matter and baryogenesis and implement them in explicit models.

Determining dynamically the symmetry breaking pattern that leads to the SM is a next step after geometric moduli stabilisation,  that is moduli stabilisation  in the open string sector.
 On SUSY breaking, study in detail all the different scenarios including phenomenological observables at reach at LHC and potential future colliders. In particular the sequestered scenarios \cite{aparicio}. An important challenge is to make consistent the large string and gravitino scale hinted by BICEP  with TeV soft terms. The experimental inputs are becoming more and more useful to  allow sculpting string scenarios that include all main observational constraints.

Regarding more calculation developments,  next order corrections to the EFT are needed. In particular quantum corrections to the K\"ahler potential  for the moduli and  matter fields. This is crucial to obtain reliable soft terms, including contributions to non-universality.  Also, compute $\alpha'$ corrections to moduli K\"ahler potential in F-theory compactifications to explicitly realise LVS in those models.
Finally it would be interesting to identify CFT duals of the AdS compactifications to provide them a proper non-perturbative definition and also a stringy understanding of these scenarios beyond EFT would be desirable.

 A key point to  keep in mind is to try identify potential observables that can  put classes of models to test. The dark radiation issue raised recently is a good example. It has  led to put strong constraints on hidden sector models that would give too large an effect but also to identify a  cosmic axion background (CAB) with potential observable effects manifested as X-ray excess in galaxy clusters \cite{cab}. Similar ideas have been put forward in the past regarding the potential of observing cosmic strings motivated by the way to end brane inflation, etc. These observable effects are very rare and difficult to identify but worth pursuing. Also, potential discoveries in the recent future by LHC, Planck and other CMB experiments, on dark matter, axions searches, etc. should give us  guidelines as to how to restrict or identify realistic string models. We may get lucky one of these days.

\section{Acknowledgements}
I would like to thank my collaborators over the past few years on the original work reported in this overview: Luis Aparicio, Ralph Blumenhagen, Cliff Burgess, Michele Cicoli, Joe Conlon, Shanta de Alwis, Anamaria Font, Rajesh Gupta, Ehsan Hatefi, Denis Klevers, Sven Krippendorf,  Anshuman Maharana, Christoph Mayrhofer, Sebastian Moster, Stefan Theisen, Roberto Valandro. Special thanks to Sven Krippendorf for comments on the manuscript.



\begin{thebibliography}{9}

\bibitem{iu}
  L.~E.~Ibanez and A.~M.~Uranga,
  \emph{String theory and particle physics: An introduction to string phenomenology,}
  Cambridge, UK: Univ. Pr. (2012).

\bibitem{witten}
  E.~Witten,
  \emph{Reflections on the fate of space-time,}
  Phys.\ Today {\bf 49N4} (1996) 24.
  
\bibitem{fqt}
  A.~Font, F.~Quevedo and S.~Theisen,
  \emph{A Comment on Continuous Spin Representations of the Poincare Group and Perturbative String Theory,}
  arXiv:1302.4771 [hep-th].

\bibitem{weinberg}
  S.~Weinberg,
  \emph{The Quantum theory of fields. Vol. 1: Foundations,}
  Cambridge, UK: Univ. Pr. (1995) 609 p

\bibitem{banksdixon}
T.~Banks and L.~J.~Dixon,
  ``Constraints on String Vacua with Space-Time Supersymmetry,''
  Nucl.\ Phys.\ B {\bf 307} (1988) 93;
C.~P.~Burgess, J.~P.~Conlon, L-Y.~Hung, C.~H.~Kom, A.~Maharana and F.~Quevedo,
  ``Continuous Global Symmetries and Hyperweak Interactions in String Compactifications,''
  JHEP {\bf 0807} (2008) 073
  [arXiv:0805.4037 [hep-th]].

\bibitem{cmp}
  G.~D.~Coughlan, W.~Fischler, E.~W.~Kolb, S.~Raby and G.~G.~Ross,
  \emph{Cosmological Problems for the Polonyi Potential,}
  Phys.\ Lett.\ B {\bf 131} (1983) 59;\\
  T.~Banks, D.~B.~Kaplan and A.~E.~Nelson,
  \emph{Cosmological implications of dynamical supersymmetry breaking,}
  Phys.\ Rev.\ D {\bf 49} (1994) 779
  [hep-ph/9308292];\\
  B.~de Carlos, J.~A.~Casas, F.~Quevedo and E.~Roulet,
  \emph{Model independent properties and cosmological implications of the dilaton and moduli sectors of 4-d strings,}
  Phys.\ Lett.\ B {\bf 318} (1993) 447
  [hep-ph/9308325].


\bibitem{gkp}
  S.~B.~Giddings, S.~Kachru and J.~Polchinski,
  \emph{Hierarchies from fluxes in string compactifications,}
  Phys.\ Rev.\ D {\bf 66} (2002) 106006
  [hep-th/0105097].

\bibitem{dk}
  M.~R.~Douglas and S.~Kachru,
  \emph{Flux compactification,}
  Rev.\ Mod.\ Phys.\  {\bf 79} (2007) 733
  [hep-th/0610102].

\bibitem{lvs}
  V.~Balasubramanian, P.~Berglund, J.~P.~Conlon and F.~Quevedo,
  \emph{Systematics of moduli stabilisation in Calabi-Yau flux compactifications,}
  JHEP {\bf 0503} (2005) 007
  [hep-th/0502058];\\
  J.~P.~Conlon, F.~Quevedo and K.~Suruliz,
  \emph{Large-volume flux compactifications: Moduli spectrum and D3/D7 soft supersymmetry breaking,}
  JHEP {\bf 0508} (2005) 007
  [hep-th/0505076].

\bibitem{kklt}
  S.~Kachru, R.~Kallosh, A.~D.~Linde and S.~P.~Trivedi,
  \emph{De Sitter vacua in string theory,}
  Phys.\ Rev.\ D {\bf 68} (2003) 046005
  [hep-th/0301240].

\bibitem{bkq}
  C.~P.~Burgess, R.~Kallosh and F.~Quevedo,
  \emph{De Sitter string vacua from supersymmetric D terms,}
  JHEP {\bf 0310} (2003) 056
  [hep-th/0309187].


\bibitem{desitter}
  S.~L.~Parameswaran and A.~Westphal,
  \emph{de Sitter string vacua from perturbative Kahler corrections and consistent D-terms,}
  JHEP {\bf 0610} (2006) 079
  [hep-th/0602253];\\
  M.~Cicoli, A.~Maharana, F.~Quevedo and C.~P.~Burgess,
  \emph{De Sitter String Vacua from Dilaton-dependent Non-perturbative Effects,}
  JHEP {\bf 1206} (2012) 011
  [arXiv:1203.1750 [hep-th]];\\
  J.~BlŒbŠck, D.~Roest and I.~Zavala,
  \emph{De Sitter Vacua from Non-perturbative Flux Compactifications,}
  arXiv:1312.5328 [hep-th].

\bibitem{wbounds}
S.~P.~de Alwis,
  \emph{Constraints on LVS Compactifications of IIB String Theory,}
  JHEP {\bf 1205} (2012) 026
  [arXiv:1202.1546 [hep-th]];
M.~Cicoli, J.~P.~Conlon, A.~Maharana and F.~Quevedo,
  \emph{A Note on the Magnitude of the Flux Superpotential,}
  JHEP {\bf 1401} (2014) 027
  [arXiv:1310.6694 [hep-th], arXiv:1310.6694].

\bibitem{stability}
  S.~de Alwis, R.~Gupta, E.~Hatefi and F.~Quevedo,
  \emph{Stability, Tunneling and Flux Changing de Sitter Transitions in the Large Volume String Scenario,}
  JHEP {\bf 1311} (2013) 179
  [arXiv:1308.1222 [hep-th], arXiv:1308.1222].


\bibitem{uber}
  C.~P.~Burgess, A.~Maharana and F.~Quevedo,
  \emph{Uber-naturalness: unexpectedly light scalars from supersymmetric extra dimensions,}
  JHEP {\bf 1105} (2011) 010
  [arXiv:1005.1199 [hep-th]].



\bibitem{planck}
  C.~P.~Burgess, M.~Cicoli and F.~Quevedo,
  \emph{String Inflation After Planck 2013,}
  JCAP {\bf 1311} (2013) 003
  [arXiv:1306.3512, arXiv:1306.3512 [hep-th]].

\bibitem{larger}
A.~R.~Liddle, A.~Mazumdar and F.~E.~Schunck,
  ``Assisted inflation,''
  Phys.\ Rev.\ D {\bf 58} (1998) 061301
  [astro-ph/9804177];
  S.~Dimopoulos, S.~Kachru, J.~McGreevy and J.~G.~Wacker,
  \emph{N-flation,}
  JCAP {\bf 0808} (2008) 003
  [hep-th/0507205];
  M.~Cicoli, K.~Dutta and A.~Maharana,
  \emph{N-flation with Hierarchically Light Axions in String Compactifications,}
  arXiv:1401.2579 [hep-th];
  E.~Silverstein and A.~Westphal,
  \emph{Monodromy in the CMB: Gravity Waves and String Inflation,}
  Phys.\ Rev.\ D {\bf 78} (2008) 106003
  [arXiv:0803.3085 [hep-th]];
  L.~McAllister, E.~Silverstein and A.~Westphal,
  \emph{Gravity Waves and Linear Inflation from Axion Monodromy,}
  Phys.\ Rev.\ D {\bf 82} (2010) 046003
  [arXiv:0808.0706 [hep-th]];
  N.~Kaloper and L.~Sorbo,
  \emph{A Natural Framework for Chaotic Inflation,}
  Phys.\ Rev.\ Lett.\  {\bf 102} (2009) 121301
  [arXiv:0811.1989 [hep-th]];
  N.~Kaloper, A.~Lawrence and L.~Sorbo,
  \emph{An Ignoble Approach to Large Field Inflation,}
  JCAP {\bf 1103}, 023 (2011)
  [arXiv:1101.0026 [hep-th]];
  A.~Avgoustidis, D.~Cremades and F.~Quevedo,
  \emph{Wilson line inflation,}
  Gen.\ Rel.\ Grav.\  {\bf 39} (2007) 1203
  [hep-th/0606031];
  A.~Avgoustidis and I.~Zavala,
  \emph{Warped Wilson Line DBI Inflation,}
  JCAP {\bf 0901} (2009) 045
  [arXiv:0810.5001 [hep-th]];
F.~Marchesano, G.~Shiu and A.~M.~Uranga,
  ``F-term Axion Monodromy Inflation,''
  arXiv:1404.3040 [hep-th];
  See also:  T.~W.~Grimm,
  ``Axion Inflation in F-theory,''
  arXiv:1404.4268 [hep-th];
  R.~Blumenhagen and E.~Plauschinn,
  ``Towards Universal Axion Inflation and Reheating in String Theory,''
  arXiv:1404.3542 [hep-th].
  

\bibitem{darkradiation}
  M.~Cicoli, J.~P.~Conlon and F.~Quevedo,
  \emph{Dark Radiation in LARGE Volume Models,}
  Phys.\ Rev.\ D {\bf 87} (2013) 043520
  [arXiv:1208.3562 [hep-ph]];\\
  T.~Higaki and F.~Takahashi,
  \emph{Dark Radiation and Dark Matter in Large Volume Compactifications,}
  JHEP {\bf 1211} (2012) 125
  [arXiv:1208.3563 [hep-ph]].


\bibitem{michele}
  M.~Cicoli, \emph{Axion-like Particles from String Compactifications,}
  arXiv:1309.6988 [hep-th].

\bibitem{mirage}
  K.~Choi, A.~Falkowski, H.~P.~Nilles and M.~Olechowski,
  \emph{Soft supersymmetry breaking in KKLT flux compactification,}
  Nucl.\ Phys.\ B {\bf 718} (2005) 113
  [hep-th/0503216].

\bibitem{soft1}
  J.~P.~Conlon and F.~Quevedo,
  \emph{Gaugino and Scalar Masses in the Landscape,}
  JHEP {\bf 0606} (2006) 029
  [hep-th/0605141];\\
  J.~P.~Conlon, S.~S.~Abdussalam, F.~Quevedo and K.~Suruliz,
  \emph{Soft SUSY Breaking Terms for Chiral Matter in IIB String Compactifications,}
  JHEP {\bf 0701} (2007) 032
  [hep-th/0610129];\\
  J.~P.~Conlon, C.~H.~Kom, K.~Suruliz, B.~C.~Allanach and F.~Quevedo,
  \emph{Sparticle Spectra and LHC Signatures for Large Volume String Compactifications,}
  JHEP {\bf 0708} (2007) 061
  [arXiv:0704.3403 [hep-ph]].


\bibitem{ibanez}
  A.~Hebecker, A.~K.~Knochel and T.~Weigand,
  \emph{A Shift Symmetry in the Higgs Sector: Experimental Hints and Stringy Realizations,}
  JHEP {\bf 1206} (2012) 093
  [arXiv:1204.2551 [hep-th]].
  L.~E.~Ibanez, F.~Marchesano, D.~Regalado and I.~Valenzuela,
  \emph{The Intermediate Scale MSSM, the Higgs Mass and F-theory Unification,}
  JHEP {\bf 1207} (2012) 195
  [arXiv:1206.2655 [hep-ph]].
  L.~E.~Ibanez and I.~Valenzuela,
  \emph{The Higgs Mass as a Signature of Heavy SUSY,}
  JHEP {\bf 1305} (2013) 064
  [arXiv:1301.5167 [hep-ph]].

\bibitem{hall}
  L.~J.~Hall and Y.~Nomura,
  \emph{Grand Unification and Intermediate Scale Supersymmetry,}
  arXiv:1312.6695 [hep-ph].

\bibitem{soft2}
  R.~Blumenhagen, J.~P.~Conlon, S.~Krippendorf, S.~Moster and F.~Quevedo,
  \emph{SUSY Breaking in Local String/F-Theory Models,}
  JHEP {\bf 0909} (2009) 007
  [arXiv:0906.3297 [hep-th]].

\bibitem{aparicio}
L.~Aparicio {\it et al.}, to appear.

\bibitem{heterotic}
  L.~B.~Anderson, J.~Gray, A.~Lukas and B.~Ovrut,
  \emph{Stabilizing the Complex Structure in Heterotic Calabi-Yau Vacua,}
  JHEP {\bf 1102} (2011) 088
  [arXiv:1010.0255 [hep-th]];
  \emph{Stabilizing All Geometric Moduli in Heterotic Calabi-Yau Vacua,}
  Phys.\ Rev.\ D {\bf 83} (2011) 106011
  [arXiv:1102.0011 [hep-th]];
  M.~Cicoli, S.~de Alwis and A.~Westphal,
  \emph{Heterotic Moduli Stabilisation,}
  JHEP {\bf 1310} (2013) 199
  [arXiv:1304.1809 [hep-th]].

\bibitem{akk}
  B.~S.~Acharya, G.~Kane and P.~Kumar,
  \emph{Compactified String Theories -- Generic Predictions for Particle Physics,}
  Int.\ J.\ Mod.\ Phys.\ A {\bf 27} (2012) 1230012
  [arXiv:1204.2795 [hep-ph]].
  
  \bibitem{bmp}
  R.~Blumenhagen, S.~Moster and E.~Plauschinn,
 \emph{Moduli Stabilisation versus Chirality for MSSM like Type IIB Orientifolds,}
  JHEP {\bf 0801} (2008) 058
  [arXiv:0711.3389 [hep-th]].
  
  \bibitem{mirror}
  J.~P.~Conlon,
  \emph{Mirror Mediation,}
  JHEP {\bf 0803} (2008) 025
  [arXiv:0710.0873 [hep-th]].

  
\bibitem{mp}
  A.~Maharana and E.~Palti,
  \emph{Models of Particle Physics from Type IIB String Theory and F-theory: A Review,}
  Int.\ J.\ Mod.\ Phys.\ A {\bf 28} (2013) 1330005
  [arXiv:1212.0555 [hep-th]].

\bibitem{aiq}
  G.~Aldazabal, L.~E.~Ibanez and F.~Quevedo,
  \emph{A $D^-$ brane alternative to the MSSM,}
  JHEP {\bf 0002} (2000) 015
  [hep-ph/0001083].


\bibitem{aiqu}
  G.~Aldazabal, L.~E.~Ibanez, F.~Quevedo and A.~M.~Uranga,
  \emph{D-branes at singularities: A Bottom up approach to the string embedding of the standard model,}
  JHEP {\bf 0008} (2000) 002
  [hep-th/0005067].

\bibitem{verlinde}
  H.~Verlinde and M.~Wijnholt,
  \emph{Building the standard model on a D3-brane,}
  JHEP {\bf 0701} (2007) 106
  [hep-th/0508089];
  M.~Buican, D.~Malyshev, D.~R.~Morrison, H.~Verlinde and M.~Wijnholt,
  \emph{D-branes at Singularities, Compactification, and Hypercharge,}
  JHEP {\bf 0701} (2007) 107
  [hep-th/0610007];
  M.~Wijnholt,
  \emph{Geometry of Particle Physics,}
  Adv.\ Theor.\ Math.\ Phys.\  {\bf 13} (2009)
  [hep-th/0703047].

\bibitem{sven}
  S.~Krippendorf, M.~J.~Dolan, A.~Maharana and F.~Quevedo,
  \emph{D-branes at Toric Singularities: Model Building, Yukawa Couplings and Flavour Physics,}
  JHEP {\bf 1006} (2010) 092
  [arXiv:1002.1790 [hep-th]];\\
  M.~J.~Dolan, S.~Krippendorf and F.~Quevedo,
  \emph{Towards a Systematic Construction of Realistic D-brane Models on a del Pezzo Singularity,}
  JHEP {\bf 1110} (2011) 024
  [arXiv:1106.6039 [hep-th]].


\bibitem{ks}
  M.~Kreuzer and H.~Skarke,
  \emph{Complete classification of reflexive polyhedra in four-dimensions,}
  Adv.\ Theor.\ Math.\ Phys.\  {\bf 4} (2002) 1209
  [hep-th/0002240];\\
  \emph{PALP: A Package for analyzing lattice polytopes with applications to toric geometry,}
  Comput.\ Phys.\ Commun.\  {\bf 157} (2004) 87
  [math/0204356 [math-sc]].

\bibitem{cmq}
  J.~P.~Conlon, A.~Maharana and F.~Quevedo,
  \emph{Towards Realistic String Vacua,}
  JHEP {\bf 0905} (2009) 109
  [arXiv:0810.5660 [hep-th]].

\bibitem{global}
  M.~Cicoli, S.~Krippendorf, C.~Mayrhofer, F.~Quevedo and R.~Valandro,
  \emph{D-Branes at del Pezzo Singularities: Global Embedding and Moduli Stabilisation,}
  JHEP {\bf 1209} (2012) 019
  [arXiv:1206.5237 [hep-th]];\\
  \emph{D3/D7 Branes at Singularities: Constraints from Global Embedding and Moduli Stabilisation,}
  JHEP {\bf 1307} (2013) 150
  [arXiv:1304.0022 [hep-th]].

\bibitem{per}
  V.~Balasubramanian, P.~Berglund, V.~Braun and I.~Garcia-Etxebarria,
  \emph{Global embeddings for branes at toric singularities,}
  JHEP {\bf 1210} (2012) 132
  [arXiv:1201.5379 [hep-th]].



\bibitem{cmv}
  M.~Cicoli, C.~Mayrhofer and R.~Valandro,
  \emph{Moduli Stabilisation for Chiral Global Models,}
  JHEP {\bf 1202} (2012) 062
  [arXiv:1110.3333 [hep-th]].


\bibitem{complexstructure}
  J.~Louis, M.~Rummel, R.~Valandro and A.~Westphal,
  \emph{Building an explicit de Sitter,}
  JHEP {\bf 1210} (2012) 163
  [arXiv:1208.3208 [hep-th]];
  D.~Martinez-Pedrera, D.~Mehta, M.~Rummel and A.~Westphal,
  \emph{Finding all flux vacua in an explicit example,}
  JHEP {\bf 1306} (2013) 110
  [arXiv:1212.4530 [hep-th]].

\bibitem{complexkahler}
  M.~Cicoli, D.~Klevers, S.~Krippendorf, C.~Mayrhofer, F.~Quevedo and R.~Valandro,
  \emph{Explicit de Sitter Flux Vacua for Global String Models with Chiral Matter,}
  arXiv:1312.0014 [hep-th].


\bibitem{cab}
  J.~P.~Conlon and M.~C.~D.~Marsh,
  \emph{Searching for a 0.1-1 keV Cosmic Axion Background,}
  Phys.\ Rev.\ Lett.\  {\bf 111} (2013) 151301
  [arXiv:1305.3603 [astro-ph.CO]];
  S.~Angus, J.~P.~Conlon, M.~C.~D.~Marsh, A.~Powell and L.~T.~Witkowski,
  \emph{Soft X-ray Excess in the Coma Cluster from a Cosmic Axion Background,}
  arXiv:1312.3947 [astro-ph.HE].


\end{thebibliography}
\end{document}